\pdfoutput=1 
\documentclass{article}
\usepackage{dcase2023_techrep,amsmath,graphicx,url,times,booktabs, tabularx, subcaption, float}

\title{Data Efficient Acoustic Scene Classification using Teacher-Informed Confusing Class Instruction}

\name{Jisheng Bai, Santi Peksi, Woon-Seng Gan }
 \name{Jin Jie Sean Yeo$^{1}\sthanks{This research is supported by the Singapore Ministry of Education, Academic Research Fund Tier 2, under research grant MOE-T2EP20221-0014.}$,
       Ee-Leng Tan$^{1}$,
       Jisheng Bai$^{2}$,
       Santi Peksi$^{1}$,
       Woon-Seng Gan$^{1}$,
       }
 \secondlinename{	  
       }
 \address{$^1$ Smart Nation TRANS Lab,\\50 Nanyang Avenue, Singapore 639798, ye0024an@e.ntu.edu.sg, \{etanel, speksi, ewsgan\}@ntu.edu.sg
 \\          
         $^2$ School of Marine Science and Technology, Northwestern Polytechnical University, Xi'an, China, \\
         baijs@mail.nwpu.edu.cn\\ 
  }

\begin{document}

\ninept
\maketitle

\begin{sloppy}

\begin{abstract}
In this technical report, we describe the SNTL-NTU team's submission for Task 1 \textit{Data-Efficient Low-Complexity Acoustic Scene Classification} of the detection and classification of acoustic scenes and events (DCASE) 2024 challenge. Three systems are introduced to tackle training splits of different sizes. For small training splits, we explored reducing the complexity of the provided baseline model by reducing the number of base channels. We introduce data augmentation in the form of \textit{mixup} to increase the diversity of training samples. For the larger training splits, we use \textit{FocusNet} to provide confusing class information to an ensemble of multiple \textit{Patchout faSt Spectrogram Transformer (PaSST)} models and baseline models trained on the original sampling rate of 44.1 kHz. We use Knowledge Distillation to distill the ensemble model to the baseline student model. Training the systems on the \textit{TAU Urban Acoustic Scene 2022 Mobile development dataset} yielded the highest average testing accuracy of (62.21, 59.82, 56.81, 53.03, 47.97)\% on split (100, 50, 25, 10, 5)\% respectively over the three systems.
\end{abstract}

\begin{keywords}
Acoustic scene analysis, Depthwise Convolutional Networks, Patchout FaSt Spectrogram Transformer (PaSST), FocusNet, Knowledge Distillation
\end{keywords}

\section{INTRODUCTION}
\label{sec:intro}
Task 1 of the DCASE 2024 challenge \cite{Schmid2024Data-EfficientChallenge} involves acoustic scene classification (ASC) for 10 scenes from 12 cities using 1-sec audio samples. This task includes data-efficient training splits (5, 10, 25, 50, 100)\% with constraints on model complexity (128 kB memory, 30 MMACs).

In this submission, three systems are presented. The first system is an adapted baseline model. The second model utilizes a Teacher-Student framework using knowledge distillation \cite{Hinton2015DistillingNetwork}. The third system uses the student from the second system as a teacher model for FocusNet \cite{Zhang2022FocusNet:Classes}.

The Baseline (BL) model \cite{Schmid2023DistillingCP-Mobile} provided is a simplified version of the CP-Mobile convolutional neural network (CNN) \cite{Schmid2022CP-JKUTransformer, Schmid2023CP-JKUCp-Mobile}. In this work, we adapt the BL model and find the optimal model complexity for small training splits. We refer to the adapted baseline as the N-Base Channel Baseline (N-BCBL), where N refers to the number of base channels. These optimizations aim to prevent over-fitting on the limited training data for smaller training splits. 

CNNs are widely recognized for their state-of-the-art results in the DCASE Task 1 challenge. Numerous CNN architectures have demonstrated significant success in acoustic scene classification (ASC) tasks, particularly on the TAU urban acoustic scene 2022 mobile dataset \cite{Heittola2020AcousticSolutions}. The Patchout FaSt Spectrogram Transformer (PaSST) \cite{Koutini2021EfficientPatchout} was shown to be an effective teacher for CNNs \cite{Schmid2022CP-JKUTransformer, Schmid2022KnowledgeClassification}, achieving the top rank in two consecutive DCASE Task 1 challenges.

Despite the state-of-the-art performance of the models, a study of the class-wise accuracy achieved by the top ranking models for DCASE Task 1 over the years reveals that classes such as \textit{street\_pedestrian} and \textit{public\_square} remain difficult to classify due to the samples sharing acoustic properties of other classes. We introduce \textit{FocusNet} \cite{Zhang2022FocusNet:Classes} to tackle this issue, by encouraging a student model to pay more attention to these confusion classes.

This report is organized as follows. In section 2, the input features, augmentation techniques used, and model complexity scaling approaches are discussed. Section 3 presents the submitted systems for the challenge, and their training methodology. Section 4 introduces the teacher-student training framework. Section 5 presents the results of our submissions on the development dataset. The report is concluded in Section 6.

\section{DATA PREPROCESSING AND AUGMENTATION}
\label{sec:data aug}
\subsection{Preprocessing}
For N-BCBL models, we use 44.1 kHz audio to compute log-mel spectrograms with 256 frequency bins. Short Time Fourier Transformation (STFT) is applied with a window size of 75 ms and a hop size of 11 ms. The model has 35K parameters with 22.6 M MACS.

For PaSST models, we use audio at a sampling rate of 44.1 kHz to compute log-mel spectrograms with 128 frequency bins. STFT is applied with a window size of 18 ms and a hop size of 7 ms. The PaSST model are pre-trained on AudioSet and fine-tuning is performed using the log-mel features described above. Pitch shifting is applied by randomly changing the maximum frequency of the Mel filter bank \cite{Koutini2021DetectionReport}.

For student models, 32-BCBL models are trained on audio resampled to 32 kHz to compute log-mel spectrograms with 256 frequency bins. Short Time Fourier Transformation (STFT) is applied with a window size of 96 ms and a hop size of 16 ms. The student models have 61K parameters with 29.4 M MACS.

We remove the random roll augmentation as we did not observe a noticeable improvement in the development set. Frequency masking with a maximum size of 48 Mel bins is applied to all models.

\subsection{Frequency-MixStyle Device Impulse Response Augmentation and Mixup}
We implement Freq-MixStyle (FMS) \cite{Schmid2022KnowledgeClassification, Kim2022DomainClassification}, which normalizes the frequency bands in a spectrogram and then denormalizes them with mixed frequency statistics of two spectrograms. FMS is randomly applied to each batch with a probability, $p_{FMS} = 0.4$. The mixing coefficient is drawn from a beta distribution parameterized by the hyperparameter $\alpha$. $\alpha = 0.3$ is used for all experiments.

The device impulse responses (DIRs) from MicIRP \footnote{http://micirp.blogspot.com} were used to augment the waveforms. DIR augmentation was applied to each batch with a probability, $p_{DIR} = 0.6$. 

Mixup \cite{Zhang2018MixUpMinimization} is designed to improve the generalization ability of deep neural networks by creating new training examples through linear interpolation of existing samples. Mixup generates synthetic samples by combining pairs of training data and their corresponding labels. The mixing coefficient is drawn from a beta distribution parameterized by the hyperparameter $\alpha$. We use $\alpha = 1$ for all experiments.

\subsection{Channel Scaling}
The BL model's complexity can be adjusted in 3 ways. Namely, the number of base channels, expansion rate and channel multiplier. In this work, we adjust the base channels while keeping the expansion rate and channel multiplier constant. By reducing the model parameters, we aim to reduce over-fitting of the model on the small training splits.

\section{SUBMITTED MODELS}
\label{sec:teacher}
\subsection{N-Base Channel Baseline Model}
To optimize the model for small training splits, the complexity was reduced by reducing the number of base channels. In a CNN, the channel dimension typically increases while the feature map size decreases. By lowering the initial number of base channels, we reduce the total number of channels throughout the network. This, in turn, decreases the number of parameters, simplifying the model and potentially improving its performance on smaller datasets.

We refer to this approach as the N-Base Channel Baseline (N-BCBL) model. The N-BCBL model is trained using mixup, FMS and DIR augmentation. In this report, we submit the 24-BCBL as one of our submission models as it yields the highest performance for small training splits. 

\subsection{Knowledge Distillation Ensemble Model}
System 2 of our submission is the Knowledge Distillation Ensemble (KD-Ensemble) model. The KD-Ensemble uses knowledge distillation to compress the knowledge of an ensemble of teacher models to a 32-BCBL student. The teacher model ensemble consists of three PaSST and three 32-BCBL models. All 32-BCBL teacher models are trained using mixup. The teacher models are trained in 3 configurations - Freq-MixStyle only, DIR augmentation only or Freq-MixStyle and DIR augmentation.

The use of N-BCBL models in the ensemble is inspired by CPJKU's use of ResNet teacher models in their ensemble, to provide more diverse predictions to the teacher ensemble. The teacher logits are ensembled by averaging the logits of all teacher models. The logits are then used in Knowledge Distillation (KD) training of a N-BCBL student.

For PaSST, the training protocol is the same as in \cite{Morocutti2023}. For the N-BCBL models, mixup is used for all samples, random roll is removed and the number of warm-up steps in the learning rate schedule are reduced to 100.

\subsection{Teacher-Focused Student Model}
System 3 of our submission is the Teacher-Focused Student (TFS) model. The student model of system 2 acts as a teacher model and is used to generate pre-computed logits of the training data. FocusNet uses the logits from the teacher model to identify classes that are difficult to classify. This acts as a form of attention, which then encourages the student to learn more discriminative features. FocusNet does not change the architecture of the models and thus does not add to the complexity of the model.

\subsection{Model Architecture}
The N-BCBL model is an adaptation of the baseline based on the CP-Mobile architecture which uses CPM blocks described in \cite{Schmid2023DistillingCP-Mobile} without Global Response Normalization (GRN) after each block. Figure \ref{fig:CPM block} shows the architecture of the CPM blocks used in the BL architecture.
\begin{figure}[t]
  \centering
  \centerline{\includegraphics[width=\columnwidth]{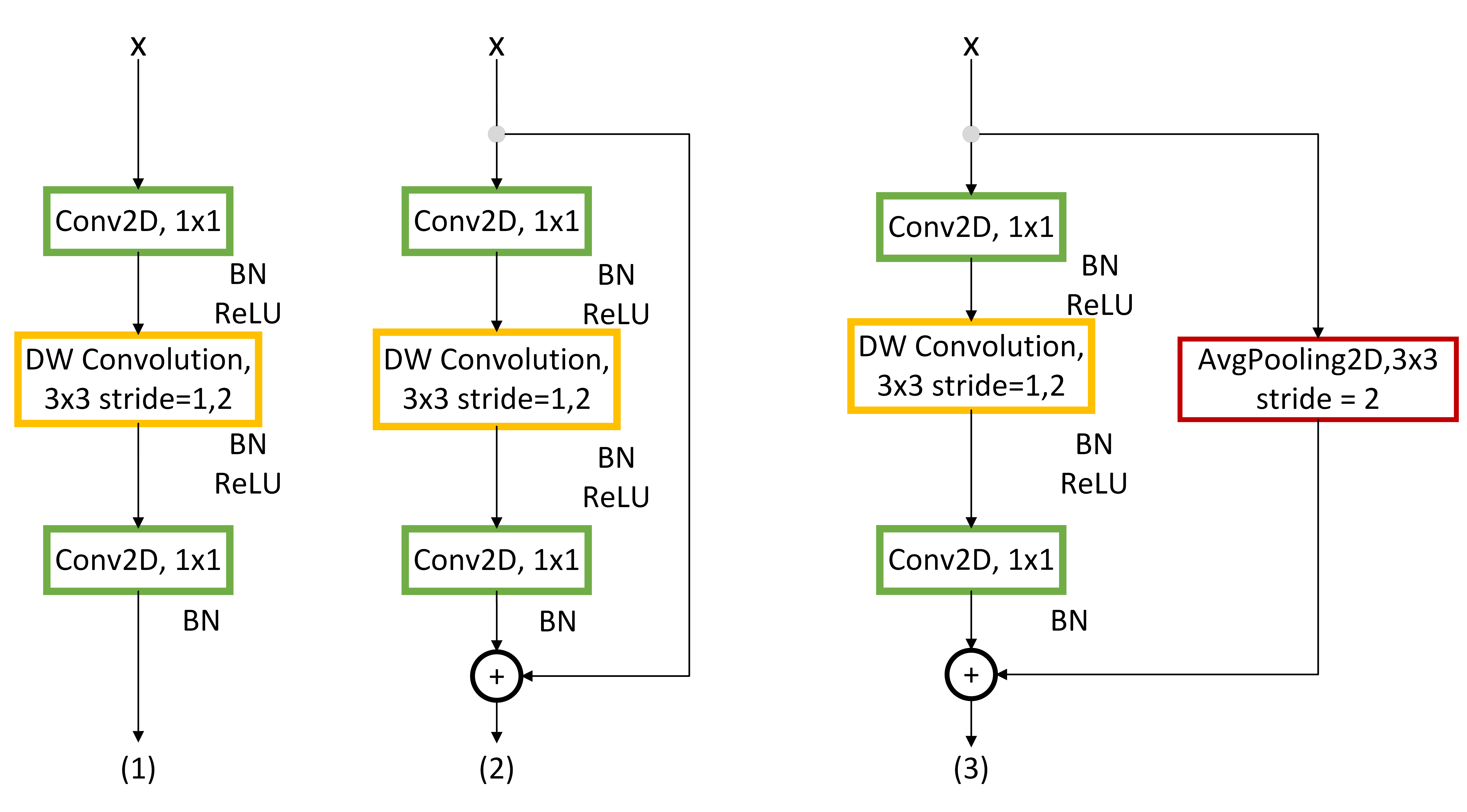}}
  \caption{CPM blocks: (1) Transition Block (input channels $\neq$ output channels
put channels), (2) Standard Block, (3) Spatial Downsampling Block}
  \label{fig:CPM block}
\end{figure}

\begin{table}[t!]
\centering
\caption{24-BCBL Architecture}
\begin{tabular}{ccc}
\toprule
\textbf{Input} & \textbf{Operator} & \textbf{Stride} \\
\midrule
256 x 89 x 1 & Conv2D@3x3, BN, ReLU & 2 x 2 \\
128 x 45 x 6 & Conv2D@3x3, BN, ReLU & 2 x 2 \\
\midrule
64 x 23 x 24 & CPM Block S & 1 x 1 \\
64 x 23 x 24 & CPM Block D & 2 x 2 \\
64 x 23 x 24 & CPM Block S & 1 x 1 \\
\midrule
64 x 12 x 24 & CPM Block T & 2 x 1 \\
32 x 12 x 40 & CPM Block S & 1 x 1 \\
\midrule
32 x 12 x 40 & CPM Block T & 1 x 1 \\
\midrule
32 x 12 x 80 & Conv2D@1x1, BN & 1 x 1 \\
32 x 12 x 10 & Avg. Pool & - \\
\bottomrule
\end{tabular}
\begin{flushleft}
\footnotesize{
\textbf{Input:} Frequency Bands x Time Frames x Channels \\
\textbf{CPM Block S/D/T:} Standard/Downsampling/Transition \\
}
\end{flushleft}
\end{table}

\begin{table}[t!]
\centering
\caption{32-BCBL Architecture}
\begin{tabular}{ccc}
\toprule
\textbf{Input} & \textbf{Operator} & \textbf{Stride} \\
\midrule
256 x 64 x 1 & Conv2D@3x3, BN, ReLU & 2 x 2 \\
128 x 33 x 8 & Conv2D@3x3, BN, ReLU & 2 x 2 \\
\midrule
64 x 17 x 32 & CPM Block S & 1 x 1 \\
64 x 17 x 32 & CPM Block D & 2 x 2 \\
64 x 17 x 32 & CPM Block S & 1 x 1 \\
\midrule
64 x 9 x 32 & CPM Block T & 2 x 1 \\
32 x 9 x 56 & CPM Block S & 1 x 1 \\
\midrule
32 x 9 x 56 & CPM Block T & 1 x 1 \\
\midrule
16 x 9 x 104 & Conv2D@1x1, BN & 1 x 1 \\
16 x 9 x 10 & Avg. Pool & - \\
\bottomrule
\end{tabular}
\begin{flushleft}
\footnotesize{
\textbf{Input:} Frequency Bands x Time Frames x Channels \\
\textbf{CPM Block S/D/T:} Standard/Downsampling/Transition \\
}
\end{flushleft}
\end{table}

\section{TEACHER-STUDENT TRAINING METHODS}
\subsection{Knowledge Distillation}
Knowledge distillation is a model compression technique where a smaller, student model is trained to replicate the behavior of a larger, more complex model teacher model. 
Equation (1) shows the distillation loss, $\mathcal{L}_{kd}$. It includes both the categorical cross entropy (CCE) loss \( H \) and the Kullback-Leibler Divergence loss \( KL \). The parameter \( \lambda \) is used to balance the contributions of the label and distillation losses. Here, \( z_s \) and \( z_t \) represent the logits from the student and teacher networks, respectively. $y$ represents training labels. \( T \) is a temperature parameter that softens the probability distributions generated by the softmax activation function \( \delta \). 
\begin{equation}
\mathcal{L}_{kd} = (1 - \lambda) H(\delta(z_s) , y) + \lambda \cdot T^2 \cdot KL(\delta(z_t / T), \delta(z_s / T)),
\end{equation}

\begin{figure}[t]
  \centering
  \centerline{\includegraphics[width=0.8\columnwidth]{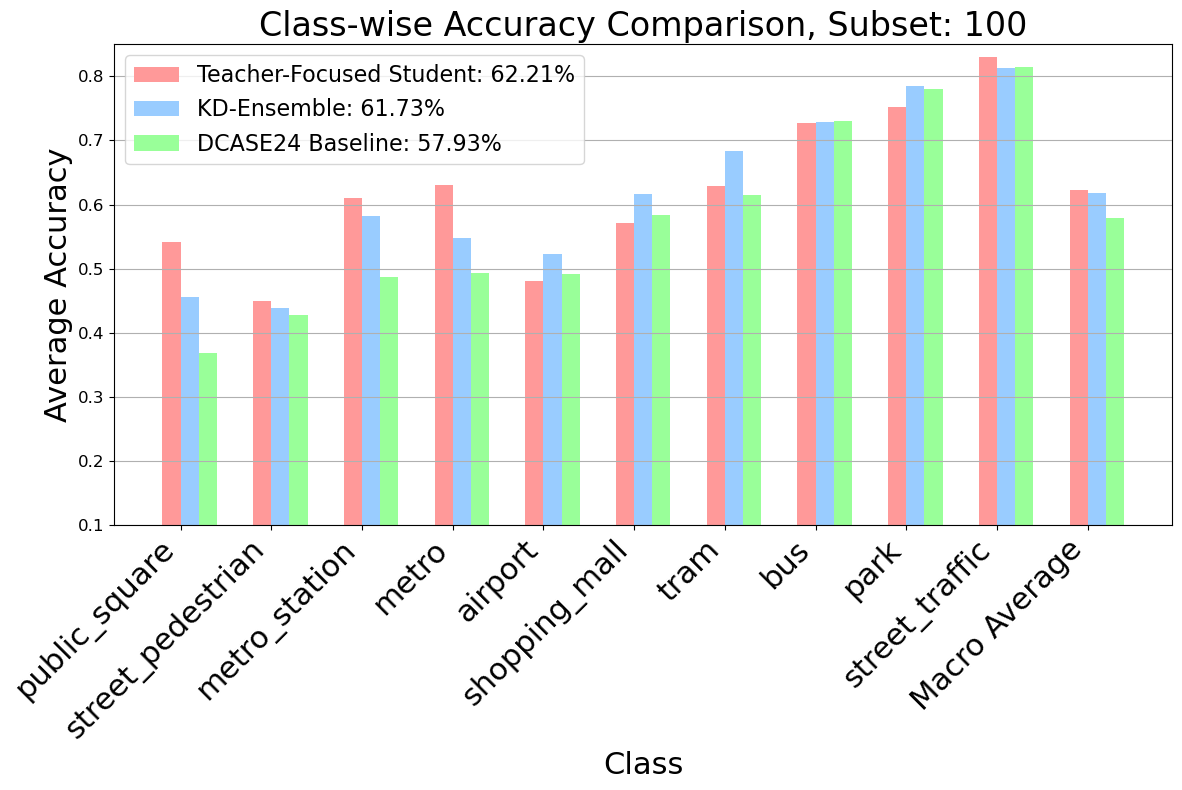}}
  \caption{Class-wise Accuracy of Student models and Baseline for the 100\% split}
  \label{fig:results_100}
\end{figure}
\subsection{FocusNet}
FocusNet aims to reduce misclassification rates by focusing on highly confused classes. The loss functions in FocusNet achieve three objectives: distinguishing between hard and easy samples, focusing on confusing classes, and preventing over-confident predictions by the student model. The overall loss function, $\mathcal{L}_{focus}$, is:

\begin{equation}
    \mathcal{L}_{focus} = \mathcal{L}_{\text{cls}} + \alpha \mathcal{R}_{\text{attention}} - \beta \mathcal{R}_{\text{entropy}},
\end{equation}
where \(\alpha\) and \(\beta\) are hyperparameters between 0 and 1. \(\mathcal{L}_{\text{cls}}\) is the classification loss, \(\mathcal{R}_{\text{attention}}\) is the attention loss, and \(\mathcal{R}_{\text{entropy}}\) is the regularization term. We set $\alpha$ and $\beta$ to 1 for all TFS models.
The classification loss first measures the difference in predictions \(d_n\), between the student and teacher model and is given by \(d = \delta(z_s ) - \delta(z_t )\). \(z_s\) and \(z_t\) are the logit values of the student and teacher models, respectively. \(d_n\) then modifies prediction \(\hat{y}_n\) as given by \(\hat{y} = \delta(d + z_s).\)

The classification loss \(\mathcal{L}_{\text{cls}}\) is computed as the cross-entropy between the ground truth label \(y\) and the modified prediction \(\hat{y}\):
\begin{equation}
    \mathcal{L}_{\text{cls}} = H(\mathbf{y}, \hat{\mathbf{y}}).
\end{equation}
Next, a multi-warm label (mwl) mask vector \(l'\) is created based on the sign of the individual teacher model's logits \(z_n^t\):

\begin{equation}
    l_n' = \begin{cases} 1, & \text{if } z_n^t > 0 \\ 0, & \text{if } z_n^t \leq 0 \end{cases}
\end{equation}

The mwl vector \(l''\) is formed by adding the mask to the ground truth label \(y_n\):

\begin{equation}
    l_n'' = \min(1, l_n' + y_n).
\end{equation}

The normalized mwl vector \(l\) is then computed as follows:

\begin{equation}
    l_n = \frac{l_n''}{\sum_{j=1}^N l_j''}.
\end{equation}

Finally, the attention loss is calculated as the cross-entropy between \(l\) and student predictions:

\begin{equation}
    \mathcal{R}_{\text{attention}} = H(l, \delta(z_s)),
\end{equation}

The regularization term \(\mathcal{R}_{\text{entropy}}\) encourages the student model to make higher entropy predictions and is computed as follows:

\begin{equation}
    \mathcal{R}_{\text{entropy}} = H(\delta(z_s),\delta(z_s)) = -\sum_{n=1}^N \delta(z_s) \log(\delta(z_s)).
\end{equation}

The regularization term is subtracted in (2) to penalize low entropy distributions, promoting higher entropy in student model predictions.

\begin{figure*}[t]
    \centering
    \includegraphics[width=\textwidth]{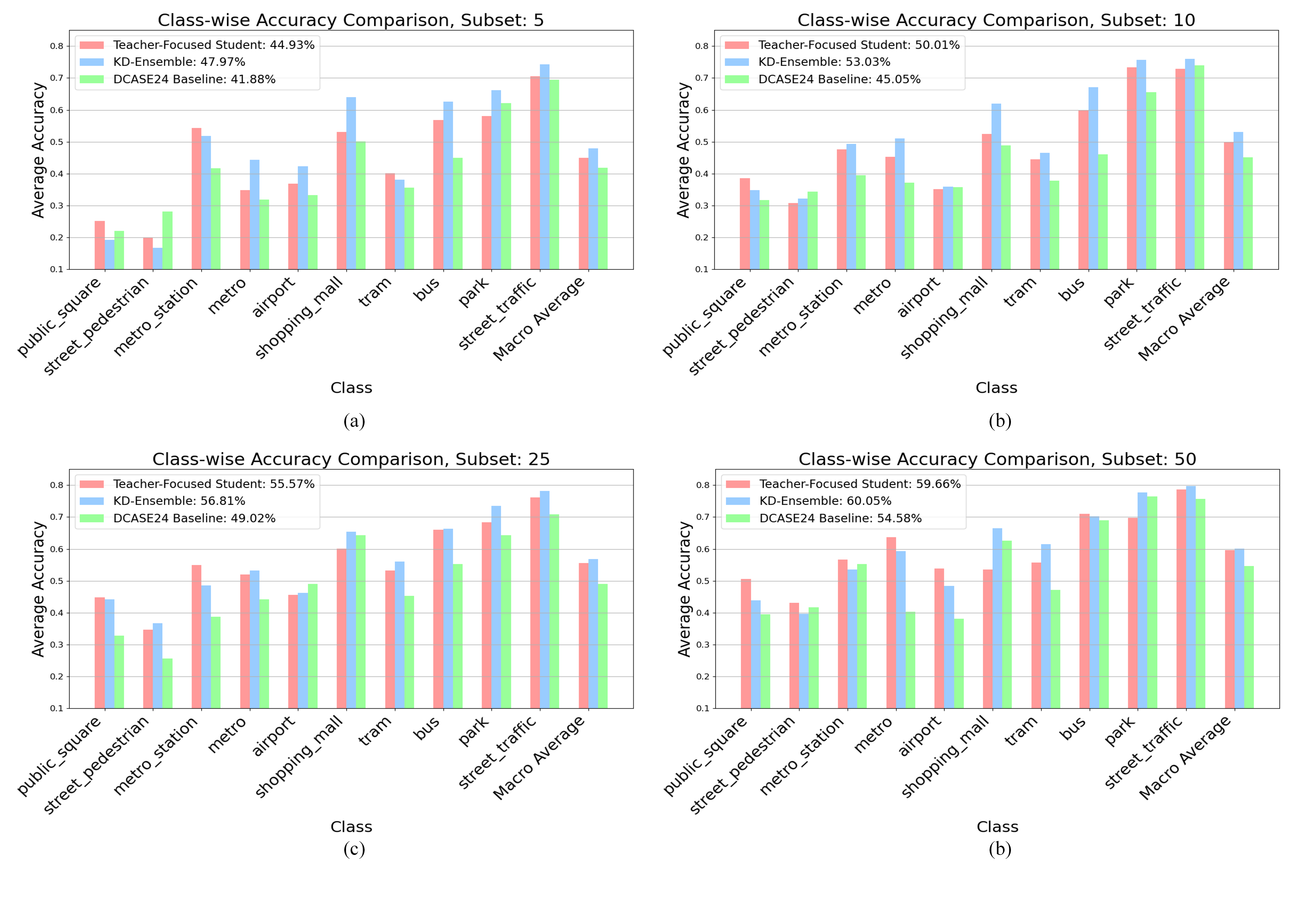}
    \caption{Class-wise Accuracy of student models and Baseline on various splits: (a) 5\%, (b) 10\%, (c) 25\%, (d) 50\%}
    \label{fig:results_all}
\end{figure*}

\section{Results}
\label{sec:results}

Table \ref{tab:accuracy_comparison} shows the accuracy results of the three systems. The TFS model achieves the highest accuracy for the 100\% split, while the KD-Ensemble achieves the highest accuracy for all other splits.

Figure \ref{fig:results_100} illustrates that the accuracy for confusing classes such as \textit{street\_pedestrian}, \textit{public\_square}, \textit{metro}, and \textit{metro\_station} improves over the KD-Ensemble and baseline models. In contrast, easy-to-classify classes like \textit{park}, \textit{bus}, and \textit{tram} show performance degradation against the KD-Ensemble and, in some cases, the baseline.

While the 24-BCBL model outperforms the baseline model (Table \ref{tab:accuracy_comparison}), both the KD-Ensemble and TFS outperform the 24-BCBL on all splits. This indicates that CNNs can benefit from the guidance of a more complex teacher, even when training data is limited.

For smaller splits, the KD-Ensemble outperforms both the TFS and 24-BCBL models. However, the ability to classify confusing classes remains effective even with small training splits as seen in Fig. \ref{fig:results_all}. An interesting research direction could be to ensemble students trained using KD and TFS models, which we leave as future work.

\begin{table}[t!]
\centering
\caption{Macro Average Accuracy Comparison}
\begin{tabular}{l|cccc}
\toprule
\textbf{Train split} & \textbf{24-BCBL} & \textbf{KD-Ensemble} & \textbf{TFS} & \textbf{BL Model} \\
\midrule
5\% & 43.68 & \textbf{47.97}  & 44.93  & 42.40  \\
10\% & 47.21 & \textbf{53.03}  & 50.01  & 45.29  \\
25\% & 53.39 & \textbf{56.81}  & 55.57  & 50.29  \\
50\% & 55.50 & \textbf{59.82}  & 59.66  & 53.19  \\
100\% & 57.59 & 61.74  & \textbf{62.21}  & 56.99 \\
\bottomrule
\end{tabular}
\label{tab:accuracy_comparison}
\end{table}

%5\% & 43.68 ± 1.15 & \textbf{47.97}  & 44.93  & 42.40 ± 0.42 \\
%10\% & 47.21 ± 0.21 & \textbf{53.03}  & 50.01  & 45.29 ± 1.01 \\
%25\% & 53.39 ± 1.59 & \textbf{56.81}  & 55.57  & 50.29 ± 0.87 \\
%50\% & 55.50 ± 1.10 & \textbf{59.82}  & 59.66  & 53.19 ± 0.68 \\
%100\% & 57.59 ± 1.45 & 61.74  & \textbf{62.21}  & 56.99 ± 1.11 \\

\section{CONCLUSION}
\label{sec:conclusion}
This report described the NTU-SNTL submission to Task 1 of the DCASE 24 challenge. Key contributions include optimizing the N-BCBL model for small datasets by reducing complexity, improving generalizability using mixup augmentation, and training a teacher-focused student by employing knowledge distillation on an ensemble of PaSST and 32-BCBL models. TFS students exhibited a trade-off in classification performance between confusing classes and easy classes. The ASC task has potential for further performance gains where TFS models are ensembled with KD-ensemble students.

\section{ACKNOWLEDGMENT}
\label{sec:ack}

This research is supported by the Ministry of Education, Singapore, under its Academic Research Fund Tier 2 (MOE-T2EP20221-0014). 

\bibliographystyle{IEEEtran}
\bibliography{refs}
%
% or list them by yourself
% \begin{thebibliography}{9}
% 
% \bibitem{dcase2016web}
%   \url{http://www.cs.tut.fi/sgn/arg/dcase2016/}.
%
% \bibitem{IEEEPDFSpec}
%   {PDF} specification for {IEEE} {X}plore$^{\textregistered}$,
%   \url{http://www.ieee.org/portal/cms_docs/pubs/confstandards/pdfs/IEEE-PDF-SpecV401.pdf}.
%
% \bibitem{PDFOpenSourceTools}
%   Creating high resolution {PDF} files for book production with 
%   open source tools, 
%   \url{http://www.grassbook.org/neteler/highres_pdf.html}.
%
% \bibitem{eWilliams1999}
% E. Williams, \emph{Fourier Acoustics: Sound Radiation and Nearfield Acoustic
%   Holography}. London, UK: Academic Press, 1999.
% 
% \bibitem{ieeecopyright}
%   \url{http://www.ieee.org/web/publications/rights/copyrightmain.html}.
%
% \bibitem{cJones2003}
% C. Jones, A. Smith, and E. Roberts, ``A sample paper in conference
%   proceedings,'' in \emph{Proc. IEEE ICASSP}, vol. II, 2003, pp. 803--806.
% 
% \bibitem{aSmith2000}
% A. Smith, C. Jones, and E. Roberts, ``A sample paper in journals,'' 
%   \emph{IEEE Trans. Signal Process.}, vol. 62, pp. 291--294, Jan. 2000.
% 
% \end{thebibliography}

\end{sloppy}
\end{document}